\documentclass[prb,superscriptaddress,showpacs,twocolumn]{revtex4}

\usepackage{amsmath,amssymb,graphicx,color,bm,soul}
\usepackage{epsfig}

\begin{document}

\title{Oscillatory dynamics of non-equilibrium dissipative exciton-polariton condensates in weak-contrast lattices}

\author{X. Ma}
\affiliation{Institute of Condensed Matter Theory and Solid State Optics, Abbe
Center of Photonics, Friedrich-Schiller-Universit\"at Jena, Max-Wien-Platz 1, 07743 Jena, Germany}

\author{I.Yu. Chestnov}
\affiliation{Department of Physics and Applied Mathematics, Vladimir State University named after A. G. and N. G. Stoletovs, 87 Gorkogo, 600000 Vladimir, Russia}

\author{M.V. Charukhchyan}
\affiliation{Department of Physics and Applied Mathematics, Vladimir State University named after A. G. and N. G. Stoletovs,  87 Gorkogo, 600000 Vladimir, Russia}

\author{A.P. Alodjants}
\affiliation{Department of Physics and Applied Mathematics, Vladimir State University named after A. G. and N. G. Stoletovs,  87 Gorkogo, 600000 Vladimir, Russia}
\affiliation{Russian Quantum Center, 100 Novaya, 143025 Skolkovo, Moscow Region, Russia}

\author{O.A. Egorov}
\affiliation{Institute of Condensed Matter Theory and Solid State Optics, Abbe
Center of Photonics, Friedrich-Schiller-Universit\"at Jena, Max-Wien-Platz 1, 07743 Jena, Germany}

\date{\today}

\begin{abstract}

We study nonlinear dynamics of exciton-polaritons in an incoherently pumped semiconductor microcavity with embedded weak-contrast lattice and coupled to an exciton reservoir.
We elucidate fundamental features of non-equilibrium exciton-polariton condensate trapped in one-dimensional periodical potential close to zero momentum (so-called `Zero-state') and to the state at the boundary of Brillouin zone (`$\pi$-state').
Within the framework of the mean-field theory, we identify different regimes of both relaxation and oscillatory dynamics of coherent exciton-polaritons governed by superpositions of Bloch eigenstates within the periodic lattice. In particular, we theoretically demonstrate stable macroscopical oscillations, akin to nonlinear Josephson oscillations, between different spectral components of a polariton condensate in the momenta-space. We elucidate a strong influence of the dissipative effects and the feedback induced by the inhomogeneity of incoherent reservoir on the dynamics of the coherent polaritons.

\end{abstract}

\maketitle

\section{Introduction}

Strongly correlated Bosonic particles   placed in the   lattice potentials represent  indispensable tool  for fundamental studies of  quantum phenomena in modern condensed matter and solid state physics~\cite{Stringari03}. In atomic optics impressive results have been obtained   due to experimental achievement of Bose-Einstein condensation (BEC) and  manipulation of  ultracold atoms   in optical lattices  with different dimensionality at low enough (tens of microKelvins) temperatures~\cite{IBloch2005}. Nowadays  atomic condensate Josephson junctions, Mott insulator -- superfluid quantum phase transition, topological phases and  quantum simulation of various Bose-Hubbard type Hamiltonians represent a  versatile platform  for quantum information technologies due to  controllable interactions between the particles~\cite{Drummod2011}.

Last decade there was a remarkable progress in this area, which was related to exciton-polaritons occurring inside a high quality semiconductor microcavity  due to the strong light-mater coupling~\cite{Weisbuch92,Houdre94,Sanvitto2012}. Extremely small effective masses of these composite bosons allow for the observation of high temperature non-equilibrium BEC with exciton-polaritons \cite{Kasprzak06,Wouters07a,Deveaud12}. Their unique optical properties, such as non-parabolic dispersion and strong nonlinearity, are derived from the hybrid light-matter nature of these elementary excitations. As a result, polariton nonlinearity overcomes conventional optical nonlinearity in both strength and, most important, in speed. This is the key to the practical applications of exciton-polaritons, pushing the limits of traditional photonics. Particularly important nonlinear effects are optical bistability \cite{Tredicucci96,Baas04,Bajoni08} and  parametric scattering of polaritons in planar semiconductor microcavities~\cite{Ciuti00,Savvidis00,Whittaker01,Ciuti03,Wouters07}. The interplay between polaritonic dispersion and strong excitonic nonlinearity allows for the self-localization effects, i.e. spontaneous formation of quantized vortices \cite{Lagoudakis08,Lagoudakis09}, the nucleation of dark solitons \cite{Amo11a,Flayac11,Hivet12,Solnyshkov12a,Flayac13} and self-confined solutions, termed cavity polariton solitons \cite{Egorov10a,Egorov10b,Egorov11a,Sich12a,Sich14a}.

In this work we study the exciton-polariton nonequilibrium condensate confined in a weak-contrast periodical potential embedded into a planar microresonator driven by   homogeneous incoherent pump [see Fig.~\ref{Fig.band_structure}(a)]. The potential patterning of a microcavity can be achieved by the variety of techniques, such as reaction ion etching~\cite{Wertz10}, mirror thickness variation~\cite{Kaitouni06}, stress application~\cite{Balili07}, metal surface deposition~\cite{Lai07,Kim13}, surface acoustical waves~\cite{Lima06}, or by optical means~\cite{Amo10c}. The idea to use a periodical potential for the diffraction management of  matter waves has been suggested for the conventional Bose-Einstein condensates in atomic systems \cite{Staliunas06,Staliunas11}. Solid-state system supporting exciton-polaritons is a very promising platform for the investigation of the spontaneous formation of coherence in periodical lattices. For instance, spontaneous build-up of in-phase (zero-state) and anti-phase ($\pi$-state) ``superfluid'' states have been observed in the array of weak periodic potential barriers in a semiconductor microcavity~\cite{Lai07}. The formation and fragmentation of the condensate into the array of wires have been observed in periodical potentials created by surface acoustic waves~\cite{Cerda10,Krizhanovskii13}. The existence of novel spatially localized states of coherent polaritonic condensates in semiconductor microcavities with fabricated periodic in-plane potentials has been predicted~\cite{Ostrovskaya13}. Note that many fundamental aspects of the nonlinear waves dynamics in lattices have been understood in purely optical dissipative systems~\cite{Egorov07d,Egorov10c}.

Apart from ultracold atoms forming atomic BEC  the lifetime of low branch polaritons is in the picosecond regime and comparable to the thermalization time for mentioned above  experiments with semiconductor microstructures. Strictly speaking, non-equilibrium open-dissipative behavior represents an important intrinsic feature of  current polaritonic solid-state systems, since polaritons are subjected to rapid radiative decay and their population maintained due to the optical pumping. The pumping creates a reservoir of hot excitons, which can relax in energy to form a polariton condensate. Even though polariton-polariton interactions are repulsive, the essentially saturable nature of exciton-polariton interactions may lead to effectively attractive nonlinearity for sufficiently low pump powers~\cite{Smirnov14}.

Various aspect of Josephson effect for the open exciton-photon system has been studied in\cite{Wouters08,Borgh10,Shelykh08}. In particular, synchronized and unsynchronized phases for a polariton system trapped in double-well potential were discussed~\cite{Wouters08,Borgh10}. The influence of Josephson effect occurring in the nonlinear regime on polarization dynamics and spontaneous spatial separation of exciton-polariton condensates with opposite polarization has been predicted~\cite{Shelykh08}. At present, the main discussion is focused on the possibility of time enhancement of coherent oscillations occurring in coupled exciton-photon  condensate system~\cite{Demirchyan14,Liew14,Dominici14}. It has been proposed that long-lived oscillations could be obtained essentially due to open character of exciton-polariton systems and interaction with pumped reservoir.

The objective of this paper is to make a fundamental theoretical  investigation of  exciton-polaritons in 1D periodical potential [Fig.~\ref{Fig.band_structure}(a)], taking into account their non-equilibrium character and coupling with exciton reservoir.

In Sec.~\ref{Ch:Model} we describe the  model of the exciton-polariton condensate formed in semiconductor microcavity  in a strong coupling regime in the presence of incoherent homogeneous optical pump. Then, we develop a mean-field model for   three coupled harmonics. This simplified model proved itself as a convenient tool for obtaining useful analytical relations for polaritonic eigenstates and band-structures of the weak-contrast polaritonic lattices in Sec.~\ref{Ch:Eigenstates}. In Sec.~\ref{Ch:Dynamics} we give a numerical analysis of different regimes of the condensate dynamics. In particular, we study dynamics of condensate oscillations in the middle (`Zero-state') and at the boundary (`$\pi$-state') of the Brillouin zone, separately. In Sec.~\ref{Ch:OriginalModel} we test the existence and stability of the obtained nonlinear oscillations in the frame of a more general open-dissipative Gross-Pitaevskii model.

\section{Theoretical Model}  \label{Ch:Model}
\begin{figure}
\includegraphics[width=0.98\linewidth]{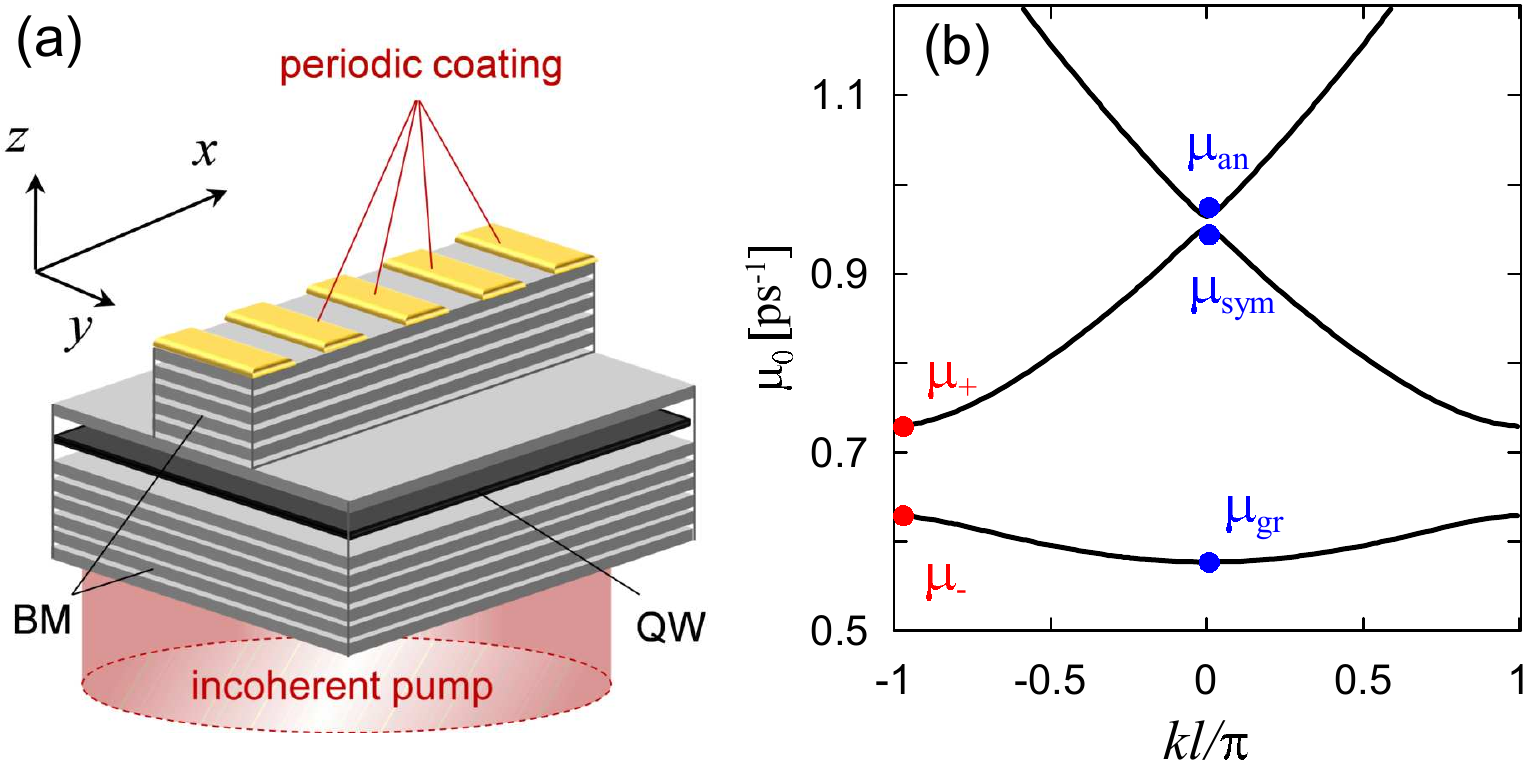}
\caption{  (Color online) (a) Sketch of the one-dimensional microcavity with a periodic coating and driven by
an incoherent optical pump. A semiconductor quantum well (QW)
is sandwiched between two Bragg mirrors (BM).  (b) Band structure of the polariton condensate in the vicinity of the condensation threshold $P_{th}=16.335$ $\mu$m$^{-2}$ps$^{-1}$. Eigenstates $\mu_{gr}$, $\mu_{an}$ and $\mu_{sym}$ represent the ground, antisymmetric and symmetric states of the polaritons with $k=0$ (`Zero-sates'), whereas $\mu_{+}$, $\mu_{-}$ are  symmetric and antisymmetric states of the polaritons with $kl=\pi$ (`$\pi$-sates'). Other parameters are: $V_0 \left/\hbar \right.=0.1$~ps${}^{-1}$, $l=2\pi/\beta=8$ $\mu$m. 
}
\label{band_structure}\label{Fig.band_structure}
\end{figure}

Using mean field theory and assuming the spontaneous formation of exciton-polariton condensate, we consider the open-dissipative Gross-Pitaevskii (GP) model that describes our incoherently pumped condensate coupled with an excitonic reservoir. Polariton order parameter $\Psi$ is described by   GP-type equation and exciton reservoir density $n$ by the rate equation~\cite{Wouters07a}:
%
%
%
\begin{align} \label{eq.polar}
i\hbar \frac{\partial \Psi}{\partial t}&= \Big[ - \frac{{{\hbar ^2}}}{{2m}}\frac{{{\partial ^2}}}{{\partial {x^2}}}+V(x)  +g_c|\Psi|^2 + i \Gamma(n)  \Big] \Psi,        \\
\frac{\partial n}{\partial t} &=-(\gamma_r + R |\Psi|^2 ) n + P_0,\label{eq.reserv}
\end{align}
where the reservoir induces the net gain and the blue shift represented by the real and imaginary parts of the term $\Gamma (n)=\frac{\hbar}{2}(R n - \gamma_c)-i g_r n $, respectively.  $R=0.01$ ps$^{-1}$~$\mu$m$^{2}$ defines the condensation rate, while $\gamma_c=0.33$ ps$^{-1}$ and $\gamma_r=0.495$ ps$^{-1}$ represent the decay rates of polaritons and reservoir excitons, respectively. Constants $g_c=6\times10^{-3}$ meV$\mu$m$^{2}$ and $g_r=2 g_c$ characterize the strengths of polariton-polariton and polariton-reservoir interactions, respectively. In Eq.~\eqref{eq.reserv} we allow  incoherent (non-resonant) homogeneous pumping, $P_0$. Here the kinetic energy of polaritons is characterized by the effective mass, $m$, which was taken as $10^{-4} m_e$ of the free electron mass.
We consider a quasi-one-dimensional polariton condensate (for instance, confined in a microwire \cite{Wertz10}) to be subjected to a periodic potential
approximated by the harmonic function $V(x)=V_0 cos(\beta x)$, where $\beta=2\pi/l$ and $l$ is the period of modulation [see Fig.~\ref{Fig.band_structure}(a)].

In the limit of zero potential ($V_0=0$) a steady-state homogeneous solution (HS) of the system (\ref{eq.polar}),(\ref{eq.reserv}) coincide with those known from the homogeneous cavity without the lattice. For the sake of generality we also allow HSs with nonzero transversal momenta $k\neq0$, which have the form of traveling waves
\begin{equation}
{\Psi _{   hs}}(x,t) = {\psi _0}{e^{ - i\,{\mu_0}t + i{k}x}},
 \label{eq:HomSol}
\end{equation}
where a condensate energy is given by $\hbar \mu_0 \equiv \hbar \mu\left( {{k},{{\left| {{\psi _0}} \right|}}} \right) = \left( {{{{\hbar ^2}} \mathord{\left/
  \right. \kern-\nulldelimiterspace} {2m}}} \right)k^2 + {g_c}{\left| {{\psi _0}} \right|^2} + {g_r}{n_{0}}$.
The HS becomes nontrivial provided that the external pump compensates for all losses overcoming the threshold value, that is ${P_{th}} = {\gamma_c \gamma _r}{ \mathord{\left/  \right. \kern-\nulldelimiterspace} {{R} }}$.
 Coherent exciton-polariton and incoherent reservoir densities are given by ${\left| {{\psi _0}} \right|^2} = {{{\kern 1pt} (P_0 - {P_{th}})} \mathord{\left/
 {} \right. \kern-\nulldelimiterspace} \gamma_c }$ and ${n_{r0}} = {\gamma_c  \mathord{\left/ {} \right. \kern-\nulldelimiterspace} { {R{\kern 1pt} } }}$, respectively~\cite{Wouters07a}.


In order to illustrate the generic properties of non-equilibrium polariton dynamics we consider a simplified approach~\cite{Staliunas06,Egorov10c,Pethick03} which takes into account only three main spatial harmonics of the weak-contrast lattice and is relevant to the description of three lowest energy bands. We expand the condensate field into a set of
spatial harmonics with the common transverse wave vector $\beta$. Index modulation
introduces a coupling between the harmonics, which lifts the degeneracy at the intersections of the dispersion curves. Then, if the solution is centered around a momentum $k$ the condensate wave function $\Psi$ and reservoir density $n$ can be approximately expressed as
\begin{subequations}
\begin{eqnarray} \label{eq.Ansaz}
\Psi (x,t) \thickapprox \left[ {A(t) + B(t){e^{ - i\beta x}} + C(t){e^{i\beta x}}} \right]{e^{ikx}},\\
n(x,t) \approx {n_0} + {n_ + }{e^{i\beta x}} + {n_ - }{e^{ - i\beta x}} , \label{eq.AnsazR}
\end{eqnarray}
\end{subequations}
where $A$, $B$, and $C$ are condensate amplitudes of spatial harmonics.
The modulation of the coherent exciton-polaritons evokes a spatial modulation of the reservoir described by the terms with $n_{\pm}$ in Eqs.~\eqref{eq.AnsazR}.

Inserting the expansions~(\ref{eq.Ansaz}) and~(\ref{eq.AnsazR}) into the Eqs.~(\ref{eq.polar}),(\ref{eq.reserv}) and collecting the terms at the same harmonics, we obtain the set of mean-field equations
\begin{widetext}
\begin{subequations} \label{eq.three}
\begin{eqnarray} 
i\hbar \frac{{\partial A}}{{\partial t}} &=& \left( { \frac{{{\hbar ^2}{k^2}}}{{2m}}+i\Gamma_0} \right)A+g_c\left( {|A{|^2} + 2|B{|^2}+ 2|C{|^2}} \right)A+ 2g_c{A^ * }BC + {0.5 V_0}\left( {B + C} \right) 
+ \Theta({n_ + }B+{n_ - }C),  \label{eq.threeA} \\ 
i\hbar \frac{{\partial B}}{{\partial t}} &=& \left( { \frac{{{\hbar ^2}{{\left( {\beta  - k} \right)}^2}}}{{2m}} + i\Gamma_0 } \right)B + g_c\left( {|B{|^2} + 2|A{|^2} + 2|C{|^2}} \right)B + g_c{C^ * }{A^2}
+{0.5 V_0}A + \Theta {n_ - }A,   \label{eq.threeB} \\ 
i\hbar \frac{{\partial C}}{{\partial t}} &=& \left( {\frac{{{\hbar ^2}{{\left( {\beta  + k} \right)}^2}}}{{2m}}+ i\Gamma_0} \right)C + {g_c}\left( {|C{|^2} + 2|A{|^2} + 2|B{|^2}} \right)C  + {g_c}{B^ * }{A^2} 
+ {0.5 V_0}A + \Theta n_ + A.   \label{eq.threeC}
\end{eqnarray}
\end{subequations}
\end{widetext}
where   parameter $\Gamma_0\equiv \frac{{\hbar }}{2}(R{n_0}-{\gamma _c})-i{g_r}{n_0}$ characterizes gain [$\rm{Re}(\Gamma_0) > 0$] or damping [$\rm{Re}(\Gamma_0)<0$] depending on the homogeneous component of the reservoir $n_0$. 
The modulation of the reservoir induces an additional coupling between spectral components with the complex amplitude  $\Theta=\left( {\frac{{i\hbar }}{2}R + {g_r}} \right)$.
These equations have to be solved self-consistently with the equations for the reservoir accounting also   the spatial modulation of the reservoir induced by the sideband components ($n_\pm$):
\begin{subequations}\label{eq.reservoir}
\begin{eqnarray}
\frac{{\partial {n_0}}}{{\partial t}} ={P_0}- \left( {{\gamma _r} + R\left( {|A{|^2} + |B{|^2} + |C{|^2}} \right)} \right){n_0} \\  \notag
- R\left( {{A^*}B + A{C^*}} \right){n_ + } - R\left( {{A^*}C + A{B^*}} \right){n_ - }, \label{eq.threeN0} \\ 
\frac{{\partial {n_ + }}}{{\partial t}} =  - \left( {{\gamma _r} + R\left( {|A|^2 + |B|^2 + |C|^2} \right)} \right){n_ + } \\
- R\left( {{A^*}C + A{B^*}} \right){n_0} - R C B^{*} n_{-}, \label{eq.threeN+} \notag \\
\frac{{\partial {n_ - }}}{{\partial t}} =  - \left( {{\gamma _r} + R\left( {|A{|^2} + |B{|^2} + |C{|^2}} \right)} \right){n_ - }  \\
- R\left( {{A^*}B + A{C^*}} \right){n_0} -R B C^{*} n_{+}. \notag \label{eq.threeN-}
\end{eqnarray}
\end{subequations}
Thus, a dissipative truncated model~[\eqref{eq.three} and \eqref{eq.reservoir}] describes nonlinear evolution of the three spatial components of the exciton-polariton condensate in weak-contrast lattices.

\section{Homogeneous steady state solution}  \label{Ch:Eigenstates}


First, within the simplified model~(\ref{eq.three}),(\ref{eq.reservoir}), we study the formation of the condensate which is just beyond the condensation threshold for homogeneous pump field ($P_0>P_{th}$). In particular, the spatial homogeneity of the pump suggests that the inhomogeneous components of the reservoir be small $n_0\simeq n_{r0} \gg n_+,n_-$. Therefore any steady-stay solution  satisfies approximately the following conservation law
\begin{align}
I_0=|A(t)|^2+|B(t)|^2+|C(t)|^2\approx  \frac{P_0}{n_{r0}R} - \frac{\gamma_r}{R}, \label{eq.Energy}
\end{align}
with the reservoir density given by its homogeneous steady-state value $n_{r0}=\gamma_c/R$.

In the vicinity of the threshold for weak-contrast lattices the band gap structure is not modified so strongly and we can neglect the influence of exciton-exciton scattering forming polariton nonlinearity. In this case the energy (frequency) of the steady-state condensate can be found in the linear limit of Eqs.~(\ref{eq.three}) by searching the solution in the form $A(t)=a{e^{-i\mu(k) t}}$, $B(t)=b{e^{-i\mu(k) t}}$ and $C(t)=c{e^{-i\mu(k) t}}$. The solution of the resulting linear eigenvalue problem provides a dispersion relation, i.e. dependence of the condensate energy $\hbar\mu_0(k)$ on the transversal momentum of   condensate $k$, see Fig.~\ref{Fig.band_structure} (b).

An important question arises: which values of the transverse momentum will be chosen by the system during the spontaneous build-up of the coherence? Recent experiments on the condensation of polaritons in periodical lattices show that the condensation appears around the high symmetry points such as   zero momentum $k=0$ point and the boundary of the Brillouin zone $k=\beta \left/ 2 \right.$ \cite{Lai07,Cerda10,Krizhanovskii13}. In the latter case the condensate is characterized by the  phase difference  between two neighbouring potential minima being equal to $\pi$; this state of the condensate being called ``$\pi$-state''. These states, namely in-phase (‘Zero-state’) and anti-phase (‘$\pi$-state’), reflect the band structure of a one dimensional polariton array  and   dynamic characteristics of meta-stable exciton-polariton condensates, see Fig.~\ref{Fig.band_structure}(b).


Below we focus on the detailed study of the condensate dynamics around the most interesting symmetry points, i.e the point of zero momentum $k=0$ (subsec. \ref{Ch:Eigenstates:ZeroStates}) and the boundary of the Brillouin zone $k=\beta \left/ 2 \right.$ (subsec. \ref{Ch:Eigenstates:PiStates}) within the approach represented above.

\subsection{Zero-states of the condensate}  \label{Ch:Eigenstates:ZeroStates}

In the limit of a vanishing potential ($V_0\rightarrow0$) the condensation build-up occurs around the energy minimum, i.e. around the ‘Zero-state’ with $k=0$ [Fig.~\ref{Fig.band_structure}(b)]. Therefore we start with the investigation of the condensate dynamics around $k=0$. Solving a corresponding eigenvalue problem one obtains analytical expressions for three eigenenergies of the condensate with the corresponding eigenvectors.

\emph{Ground state:} The ground eigenstate has minimal energy (or frequency $\mu$) which is given by
\begin{align}
{\hbar\mu _{gr}} = {g_r}{n_{r0}} + 2E_0 - \sqrt {{4E_{0} ^2} + {V_0^2}/{2}}, \label{eq.EigenVal1}
\end{align}
where we introduce characteristic energy $E_{0}= {\hbar^2 }{\beta ^2} /8m $ representing the ``kinetic energy'' of condensate at the boundary of the Brillouin zone, where $k= \beta /2$.
The corresponding eigenvector is symmetric against inversion of the momentum sign ($b=c$) and has a strong zero-momentum-component ($|a|>|b|,|c|$):
\begin{align} \label{eq.EigenVec1}
\left. {\left( {\begin{array}{*{20}{c}}
a\\
b\\
c
\end{array}} \right.} \right) = \frac{{\sqrt {{I_0}} }}{{\sqrt {2{G}\left( {{G} + {4E_0}\left/{V_0}\right.} \right)} }}\left. {\left( {\begin{array}{*{20}{c}}
{{G} + 4E_0/V_0}\\
{ - 1}\\
{ - 1}
\end{array}} \right.} \right) ,
\end{align}
where   coefficient ${G} \equiv \sqrt[{}]{{{{\left( {{{4E_0} \mathord{\left/ {} \right. \kern-\nulldelimiterspace} {{V_0}}}} \right)}^2} + 2}}$. The normalization of the eigenvector's length is given by   Eq.~(\ref{eq.Energy}).

\emph{Anti-symmetric state:} The next excited eigenstate of the condensate is antisymmetric in respect to the momentum sign inversion ($b=-c$):
\begin{align}
{\hbar\mu _{an}} = {g_r}{n_{r0}} + 4E_0, \label{eq.EigenVal2}
\end{align}
\begin{align}
\left. {\left( {\begin{array}{*{20}{c}}
a\\
b\\
c
\end{array}} \right.} \right) = \left. {\sqrt {\frac{{{I_0}}}{2}} \left( {\begin{array}{*{20}{c}}
0\\
{ + 1}\\
{ - 1}
\end{array}} \right.} \right). \label{eq.EigenVec2}
\end{align}
This state has a perfect $\rm{sin}$-like shape and is spatially shifted at the quarter of lattice period.

\emph{Symmetric state:} The third eigenstate has a symmetric shape in respect to the momentum sign inversion ($b=c$):
\begin{align}
{\hbar\mu _{sym}} = {g_r}{n_{r0}} + 2E_0 + \sqrt {{ 4E_0 ^2} + {V_0^2}/{2}}, \label{eq.EigenVal3}
\end{align}
\begin{align} \label{eq.EigenVec3}
\left. {\left( {\begin{array}{*{20}{c}}
a\\
b\\
c
\end{array}} \right.} \right) = \frac{{\sqrt {{I_0}} }}{{\sqrt {2{G}\left( {{G} - 4E_0 \left/ V_0 \right.} \right)} }}\left. {\left( {\begin{array}{*{20}{c}}
{{G} - 4E_{0}/V_0}\\
{ + 1}\\
{ + 1}
\end{array}} \right.} \right).
\end{align}
We note that unlike  the ground-state~(\ref{eq.EigenVal1},\ref{eq.EigenVec1}) the central component of the symmetric state converges to zero ($a\rightarrow0$) for a vanishing potential ($V_0\rightarrow0$).

\subsection{$\pi$-states of the condensate}  \label{Ch:Eigenstates:PiStates}

Let us examine condensate around the boundaries of the Brillouin zone for $k= \beta \left/2 \right.$ \cite{Lai07,Cerda10,Krizhanovskii13} [Fig.~\ref{Fig.band_structure}(b)].  The corresponding eigenvalue problem provides a cubic polynomial for the eigenvalue. Its exact analytical solution is very cumbersome. Without the loss of   generality for our qualitative analysis we consider only two eigenstates with minimal eigenfrequencies. In the limit of weak-contrast lattices one can neglect the $C$ component since its frequency substantially overcomes the frequencies of   components $A$ and $B$. In this approximation one finds the two eigenstates with smallest eigenfrequencies:
\begin{align}
{\hbar\mu _{\pm}} = {g_r}{n_{r0}} + E_0 \pm {V_0}\left/{2}\right., \label{eq.EigenVal4}
\end{align}
\begin{align}
\left. {\left( {\begin{array}{*{20}{c}}
a\\
b\\
c
\end{array}} \right.} \right) \approx \left. {\sqrt {\frac{{{I_0}}}{2}} \left( {\begin{array}{*{20}{c}}
+1\\
{ \pm 1}\\
{ 0}
\end{array}} \right.} \right). \label{eq.EigenVec4}
\end{align}
The sign ``$+$'' (``$-$'') in Eqs.~(\ref{eq.EigenVal4}), (\ref{eq.EigenVec4}) represents   symmetric (antisymmetric) eigenstates relative to the central momentum $k=\beta \left/ 2 \right.$. In the real space both states have a $cos$-like shape but, unlike the previous cases, their periods are twice larger than the period of the lattice.

As we will see below the presence of the eigenstates ~(\ref{eq.EigenVal4}),(\ref{eq.EigenVec4}) results in the persistent oscillation of the condensate which are similar to Josephson oscillations~\cite{Lagoudakis10}.

\section{Oscillation and Relaxation Dynamics of the Condensate}  \label{Ch:Dynamics}

Superfluid properties of the exciton-polariton condensate in the periodic potential is significantly modified, taking into account non-equilibrium processes\cite{Smirnov14}. Nonlinear dynamics of the coherent exciton-polaritons in the weak-contrast lattice is imprinted mostly by the interference and nonlinear mixing between the condensate eigenstates discussed above. The study of their stability and relaxation dynamics towards the lower energetic levels deserves a particular   interest. The dynamics of the reservoir and its saturation play an important role here. In the limit of the homogeneous reservoir ($n_0\gg n_{+},n_{-}$) the system possesses an analog of the particles conservation law~\eqref{eq.Energy}, which mimics some aspects of the conservative systems. The spatial inhomogeneity of the reservoir induces additional mechanisms of the selective mode relaxation similar to those known from   laser physics as \emph{spatial hole burning}.
In the following subsections (\ref{Ch:Dynamics:ZeroStates} and \ref{Ch:Dynamics:PiStates}) we separately study the condensate dynamics around the $k=0$ and $k=\beta \left/ 2 \right.$ momenta.

\subsection{Zero-states of the condensate}  \label{Ch:Dynamics:ZeroStates}

\begin{figure}
\includegraphics[width=1\linewidth]{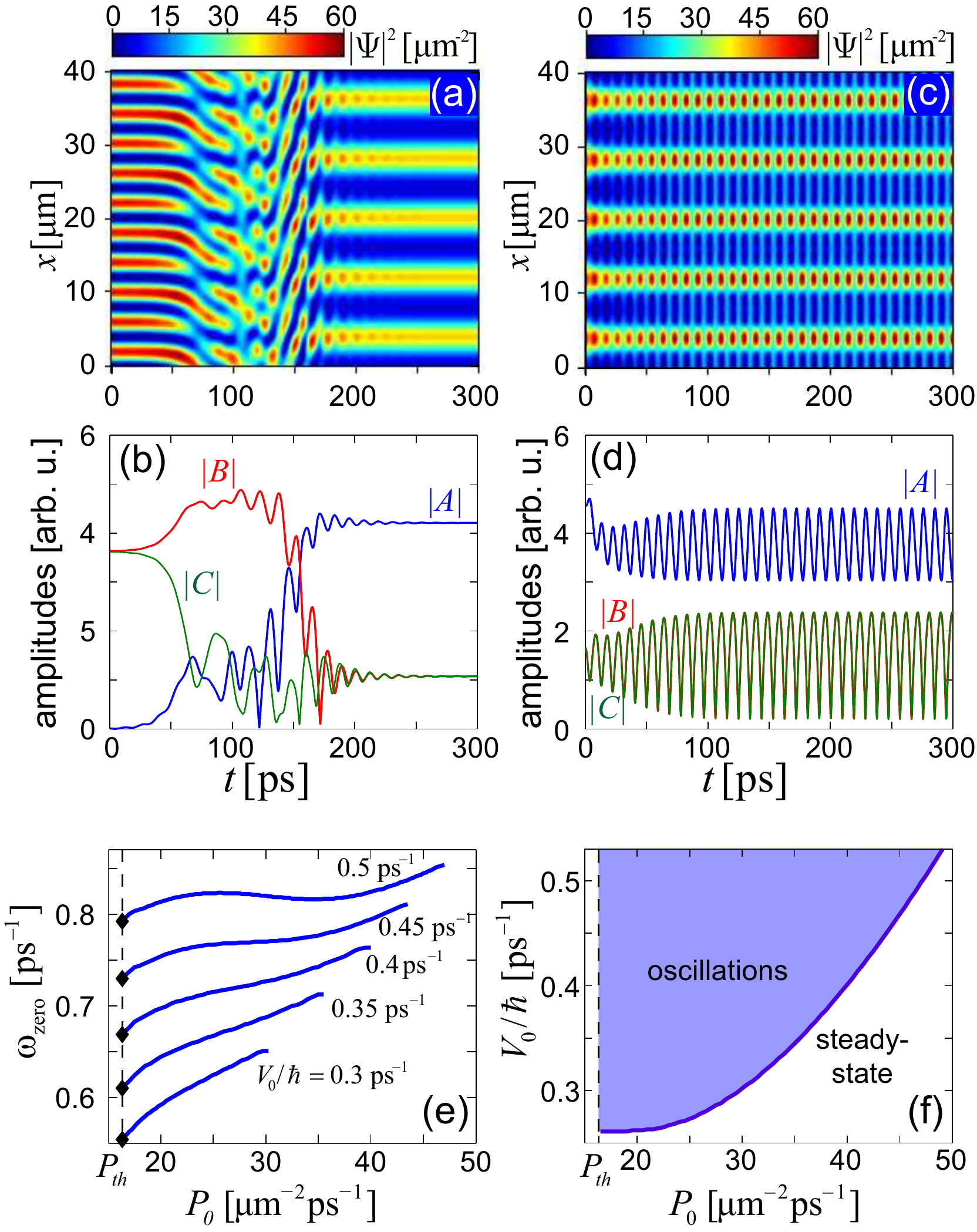}
\caption{ (Color online) Relaxation dynamics of ‘Zero-state’ polaritons for   pumping $P_0=25$~{$\mu$m$^{-2}$ps$^{-1}$} and different potential contrasts: (a,b) for $V_0 \left/ \hbar \right.=0.2 $  {ps$^{-1}$} and (c,d) for $V_0 \left/ \hbar \right.=0.35$  {ps$^{-1}$}. The figures (a,c) show   spatial profiles of   condensate density $|\Psi|^2$, whereas the figures (b,d) represent the amplitudes of the spatial harmonics ($|A|$,$|B|$,$|C|$). The initial state of the condensate is taken in the form of
antisymmetric state (\ref{eq.EigenVec2}) for (a,b) and in the form of ground
state (\ref{eq.EigenVec1}) for (b,d), respectively.
(e) Oscillation frequencies vs. the pumping rate $P_0$ for different values of   modulation depths $V_0$. Dark diamonds depict the analytical results given by Eq.~(\ref{eq:Omega_ZeroSt}). (f) Phase boundary of pump $P_0$ versus modulation depth $V_0/\hbar$ characterizing domain of persistent oscillation existence (shaded area). Vertical dashed line determines threshold value $P_{th}=16.335$~$\mu$m$^{-2}$ps$^{-1}$ of the pump.
The lattice period is the same as in Fig.~\ref{Fig.band_structure}. }
\label{fig:Dynam_ZeroSt}
\end{figure}

To illustrate the relaxation dynamics in weak-contrast lattices we properly prepared the initial distribution of exciton-polaritons in the form of the antisymmetric eigenstate given by Eqs.~(\ref{eq.EigenVal2}),~(\ref{eq.EigenVec2}). Experimentally it can be realized by launching   two coherent optical beams with a properly chosen phase difference and tilted in opposite directions. The initial state of the reservoir was taken to be spatially homogeneous, i.e. $n_0 = n_{r0}$ and $n_+ = n_- = 0$. We also took into account the initial small-amplitude noise, that has been added to the initial condensate profile. This noise breaks the symmetry of the initial eigenstate triggering the relaxation dynamics. Figures~\ref{fig:Dynam_ZeroSt} (a) and (b) clearly demonstrate dynamics and eventual relaxation to the ground states [given by Eqs.~(\ref{eq.EigenVal1}),~(\ref{eq.EigenVec1})]. In the ground state   coherent polaritons gather around the minima of the periodical potential.
%


We note that this relaxation dynamics is induced by the spatial modulation of the reservoir. Indeed, in the frame of the simplified model~[\eqref{eq.three} and \eqref{eq.reservoir}], all three polariton eigenstates experience a uniform net gain [$Re(\Gamma_0)$] provided that the reservoir modulation is negligible, i.e. $n_{\pm}=0$. Apparently the reservoir modulation ($n_{\pm}\neq0$) breaks this symmetry in favour of the ground polaritonic state. That is why the ground state becomes eventually much more populated.

This simple relaxation dynamics changes substantially for lattices with the stronger potential contrasts $V_0$. Indeed, the ground state described by Eqs.~(\ref{eq.EigenVal1},\ref{eq.EigenVec1}) becomes unstable and leads the system to the regime of persistent oscillations [Fig.~\ref{fig:Dynam_ZeroSt} (c),(d)].
The oscillations appear due to the temporal beating between the ground state and the excited symmetric mode population [Eqs.~(\ref{eq.EigenVal3},\ref{eq.EigenVec3})], cf. Fig.~\ref{fig:Dynam_ZeroSt}(b). Thus the oscillation frequency $\omega_{zero}$ can be approximated as the difference in their eigenfrequencies $\mu_{sym}$ and $\mu_{gr}$, at least, in the vicinity of the condensation threshold:
\begin{align}
\omega_{zero}\approx\sqrt{(4E_0/\hbar)^2 + 2(V_0/\hbar)^2 }. \label{eq:Omega_ZeroSt}
\end{align}
Our extensive numerical analysis of the model~\eqref{eq.three}, \eqref{eq.reservoir} shows that the Eq.~(\ref{eq:Omega_ZeroSt}) predicts accurately the oscillations frequency in the vicinity of the condensation threshold ($P_0\approx P_{th}$) where the condensate amplitudes and reservoir modulations are small [Fig.~\ref{fig:Dynam_ZeroSt}(e)]. For a stronger pumping the influence of the nonlinear effects and reservoir inhomogeneities become non-negligible. As a consequence the oscillations frequency increases with the pump amplitude $P_0$ until the critical value where the ground mode of the condensate state stabilizes again and the oscillations disappear. This threshold of the oscillations appearance is determined both by potential depth $V_0$ and pumping rate $P_0$. The domain boundary for the regime of persistent oscillations is shown in Fig.~\ref{fig:Dynam_ZeroSt}(f).


The growth of another eigenmode  against the condensate ground state can   be explained again by the selective saturation of the gain. Due to the spatial dependency of the reservoir density the gain is saturated mostly in the vicinity of the condensate maxima, i.e. in the minima of the periodical potential. It allows another mode, whose peak fields are localized near the maxima of periodical potential [i.e. the symmetric mode given by~(\ref{eq.EigenVal3},\ref{eq.EigenVec3})], the opportunity to grow as well. This phenomenon is known from laser physics as \emph{spatial hole burning}. Similar beating dynamics between two spatial modes of a weak-contrast lattice has been predicted for the optical parametric oscillator \cite{Egorov07d}.

\subsection{$\pi$-states of the condensate}  \label{Ch:Dynamics:PiStates}

\begin{figure}
\includegraphics[width=1\linewidth]{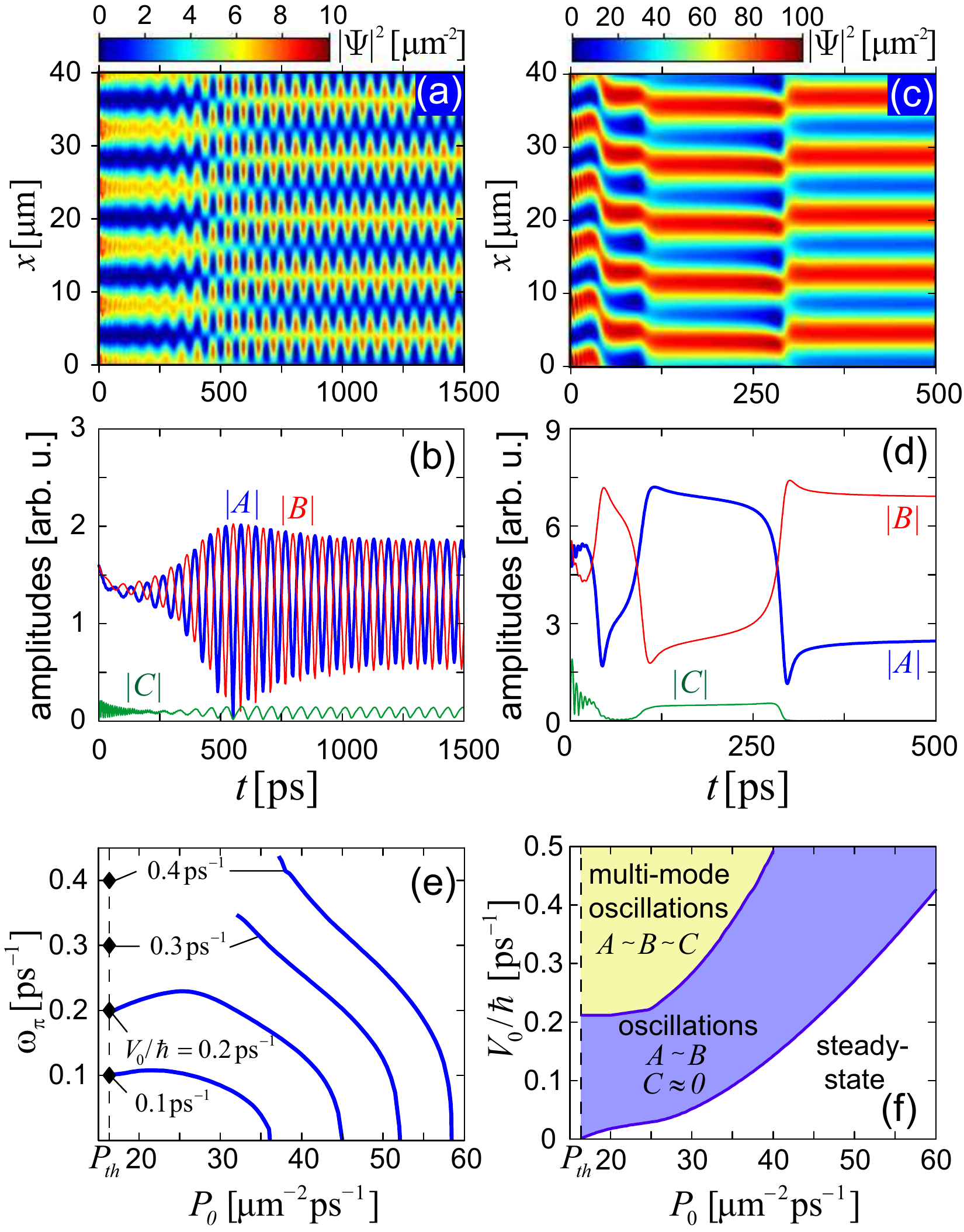}
\caption{ (Color online) Relaxation dynamics of ‘$\pi$-states’ polaritons for the potential contrasts $V_0 \left/ \hbar \right.=0.1$ {ps$^{-1}$} and different pumping rates: (a,b) for $P_0=18$ {$\mu$m$^{-2}$ps$^{-1}$} and (c,d) for $P_0=36$ {$\mu$m$^{-2}$ps$^{-1}$}. Figures (a,c) show the spatial profiles of the condensate density $|\Psi|^2$, whereas the figures (b,d) represent the amplitudes of the spatial harmonics ($|A|$,$|B|$,$|C|$).
(e) Oscillation frequencies vs. the pumping rate $P_0$ for different values of the modulation depths $V_0$. Dark diamonds depict the analytical results given by Eq.~(\ref{eq:Omega_PiSt}). (f) Phase boundary of the pump $P_0$ versus modulation depth $V_0/\hbar$. 
The lattice period is the same as in Fig.~\ref{Fig.band_structure}.}
\label{fig:Dynam_PiSt}
\end{figure}

Nonlinear relaxation dynamics of the polariton condensate can hardly be understood without considering    $\pi$-states of the condensate in a periodical potential, i.e. the states at the boundary of the Brillouin zone. Two tilted coherent beams with an appropriate photon energy (\ref{eq.EigenVal4}),~(\ref{eq.EigenVec4}) can create required spatial harmonics with the momenta $k=+\beta \left/ 2 \right.$ and $k=-\beta \left/ 2 \right.$. Our numerical analysis of the model~\eqref{eq.three},\eqref{eq.reservoir} shows that both   symmetric ($\mu_+$) and antisymmetric ($\mu_-$) eigenstates undergo instability and develop into dynamically stable persistent oscillations [Fig.~\ref{fig:Dynam_PiSt} (a),(b)]. The maxima of the condensate density oscillate around the bottoms of the periodical potential. Moreover the oscillations between two nearest neighboring potential sides are in phase.
%

Similar to the dynamics considered above~(\ref{eq:Omega_ZeroSt}) the oscillations between two $\pi$-states, namely,   symmetric $\mu_+$ and antisymmetric $\mu_-$ ones, exist even in the vicinity of the condensation threshold where the oscillation frequency can be approximated by a simple expression
\begin{align}
\omega_{\pi}\approx \mu_+ - \mu_- =V_0\left/ \hbar \right. . \label{eq:Omega_PiSt}
\end{align}
Thus, coherent polaritons periodically change their collective momenta between two values $k=+\beta \left/ 2 \right.$ and $k=-\beta \left/ 2 \right.$ which are represented in the model~\eqref{eq.three},\eqref{eq.reservoir} by the spatial harmonics $A$ and $B$, respectively [Fig.~\ref{fig:Dynam_PiSt}(b)]. This behaviour is similar to   Josephson oscillations of the polaritons in the double well potential well \cite{Lagoudakis10}. However, unlike conventional Josephson dynamics, the oscillations being considered here appear in the $k$-space, i.e. between two components with different momenta and, therefore, can be observed in the far-field measurements.


Alternatively the frequency of oscillations $\omega_{\pi}$ can be found analytically in the limit of very weak lattice contrasts ($V_0\rightarrow0$) where the spatial modulation of the reservoir can be neglected ($n_{\pm} = 0$) against the main component $n_0=\gamma_c/R$ created by the homogeneous pump. Indeed, by setting $C=0$ and $n_{\pm} = 0$ in \eqref{eq.three},\eqref{eq.reservoir}, one reduces the problem to the standard system of two coupled nonlinear equations \cite{Jensen82}. In this case, supposing that during oscillations components $A$ and $B$ completely exchange their populations,  one can obtain
\begin{align}\label{freq.pi.}
\omega_{\pi} = \frac{\pi V_0}{2\hbar F\left( \frac{\pi}{2}\left|m \right.\right)},
\end{align}
where $F\left( x \left| m\right. \right)$ is elliptic integral of the first order with $m=\frac{I_0^2g^2_c}{4V_0^2}$. The total density of the condensate $I_0=\left| A \right|^2 + \left| B \right|^2$ can be approximated by the Eq.~\ref{eq.Energy}. In the vicinity of the condensation threshold $m \ll 1$ and, thus, the frequency of $\pi$-oscillations approaches
\begin{align}\label{freq.pi.approx}
\omega_{\pi} \simeq \frac{ V_0}{\hbar} \left(1+ \frac{1}{16} \frac{I_0^2 g_c^2}{V_0^2}  \right)^{-1}.
\end{align}
We note that for $P_0 \simeq P_{th}$ the oscillations become linear, i.e. $I_0 \simeq 0$, and Eq.~\eqref{freq.pi.approx} takes the form Eq.~\eqref{eq:Omega_PiSt}.
In fact, the approach given by the Eqs.~\eqref{freq.pi.} and \eqref{freq.pi.approx} is valid in the vicinity of threshold, since, unlike the conventional case of Josephson oscillations, the system is governed by the saturation dynamics of the reservoir rather than by the Kerr-like nonlinearity.

For a stronger pumping   nonlinear effects change the character and the periods of oscillations drastically. The mean frequency ($\omega_{\pi}$) decreases along with the increase of pump $P$. Simultaneously, the form of oscillations becomes less regular, and the dynamics transforms eventually into a quasi-periodical regime. For even stronger pumping ($P_0>36$ {$\mu$m$^{-2}$ps$^{-1}$} for $V_0 / \hbar =0.1$~ps$^{-1}$) the solution becomes steady-state [Fig.~\ref{fig:Dynam_PiSt}(c),(d)].

In general, the character of   $\pi$-state  oscillations and their frequencies are sensitive to the system parameters, as it has been summarized in  Fig.~\ref{fig:Dynam_PiSt}(e). For the lattices with very weak modulation contrasts ($V_0/\hbar<0.21~$ps$^{-1}$)   persistent oscillations exist in a pump interval from the condensation threshold ($P_0\approx P_{th}$) to the critical value where the oscillation frequency approaches zero. The existence domain of the persistent oscillations are presented in   Fig.~\ref{fig:Dynam_PiSt}(f).
For   stronger modulation depths ($V_0/\hbar > 0.21~$ps$^{-1}$), however, the threshold of $\pi$-oscillations shifts to the larger values of the pumping rates and, therefore, the periodical energy exchange between $k=+\beta \left/ 2 \right.$ and $k=-\beta \left/ 2 \right.$ components are impossible in the linear limit.
Notably, $\pi$-oscillation regime discussed above is violated in the upper (depicted by ``multi-mode oscillations'' ) region in Fig.~\ref{fig:Dynam_PiSt}(f). Being in this regime the exciton-polariton system exhibits oscillatory behavior for which the population of mode $\left|C \right|^2$ becomes comparable with $\left|A \right|^2$ and $\left|B \right|^2$.

\section{Nonlinear dynamics within the Gross-Pitaevskii dissipative model}  \label{Ch:OriginalModel}

In the previous sections we discussed nonlinear dynamics in the frame of a simplified model~(\ref{eq.threeA}-\ref{eq.threeN-}). However, to obtain a more general picture of the relaxation and oscillations between all available spatial components, one has to use the original model represented by Eqs.~\eqref{eq.polar},\eqref{eq.reserv}. In this section we prove the existence and stability of both types of  oscillatory solutions obtained above by means of  the extensive numerical analysis within the frame of the open-dissipative Gross-Pitaevskii model with the periodical potential.
\begin{figure}
\includegraphics[width=1.0\linewidth]{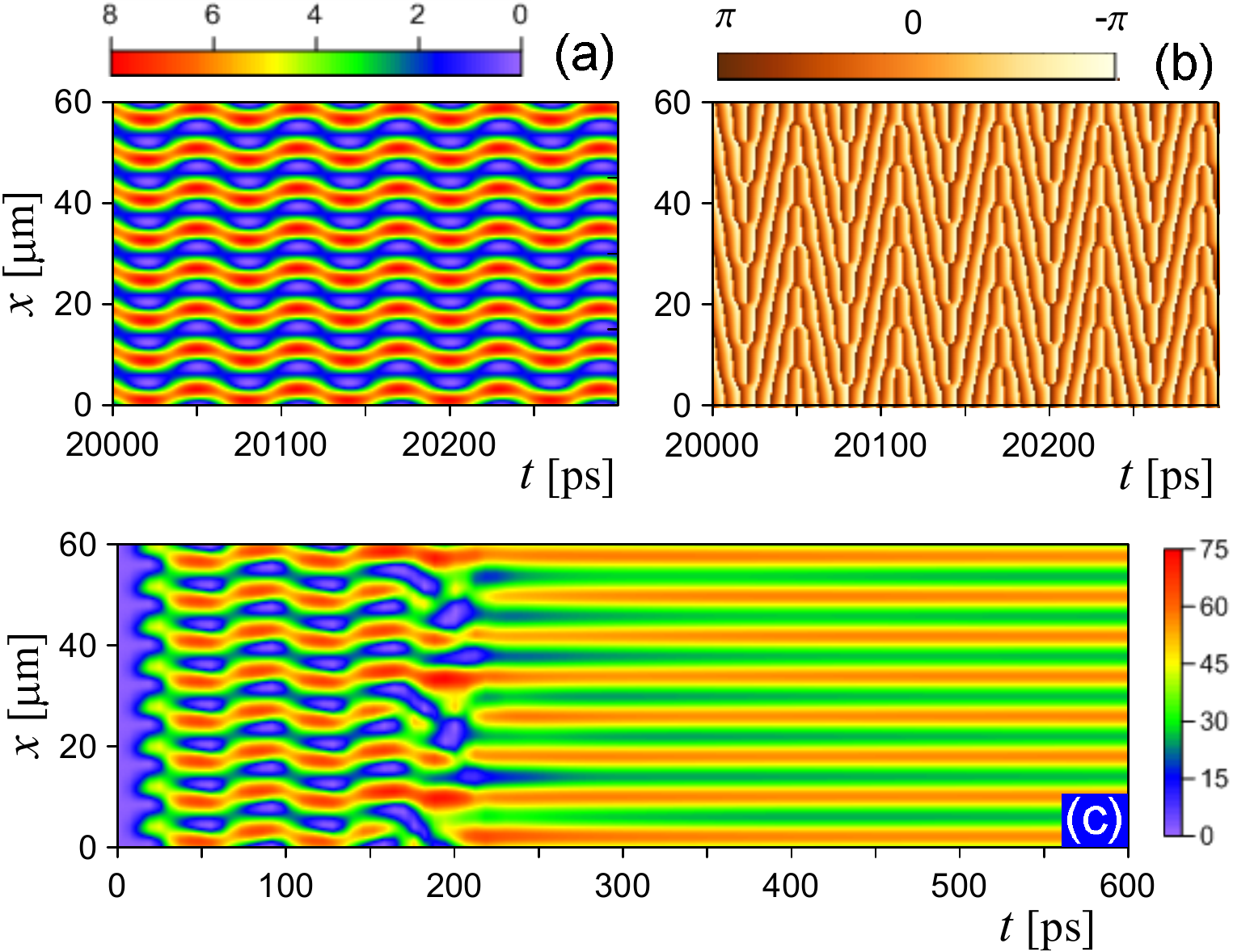}
\caption{ (Color online) Oscillation dynamics of the condensate calculated within the original model~(\ref{eq.polar}),(\ref{eq.reserv}). (a) The profile of the condensate density $|\Psi|^2$ of   oscillating $\pi$-states for the potential contrasts $V_0 \left/ \hbar \right.=0.1 $ {ps$^{-1}$} and $P_0=18$  {$\mu$m$^{-2}$ps$^{-1}$}. (b) The phase profile of the dynamics shown in (a). (c) $|\Psi|^2$ profile of the meta-stable oscillations for the pump $P_0=30$  {$\mu$m$^{-2}$ps$^{-1}$}.}
\label{fig:Dynam_GL1}
\end{figure}

First, we applied the appropriately tilted optical beam and created the initial condensate with the central momentum $k\approx\beta \left/ 2 \right.$. As a result we observed   persistent oscillations over a long time exceeding, being at least several tenths of nanoseconds  [Fig.~\ref{fig:Dynam_GL1}(a)]. The evolution of the phase profile shows that coherent polaritons really change periodically their collective momenta between two values $k=+\beta \left/ 2 \right.$ and $k=-\beta \left/ 2 \right.$ [Fig.~\ref{fig:Dynam_GL1}~(b)]. This is in perfect agreement with our analytical predictions given above [see Eq.~\eqref{freq.pi.approx} and   Fig.~\ref{fig:Dynam_PiSt}(a) and (e)]. To check the stability of these oscillations we added some additional noise to the initial condensate distribution. Note that these oscillations are very robust and are insensible to the particular value of the amplitude of the initial tilted beam. We found that both quantitative and qualitative agreements between models are held better for   smaller modulation depths of the potential $V_0$ and for smaller pumping rates $P_0$. For a stronger pumping the oscillations can become dynamically unstable and the system jumps spontaneously to the ground steady state with the minimal condensate energy [Fig.~\ref{fig:Dynam_GL1}(c)].

\begin{figure}
\includegraphics[width=1.0\linewidth]{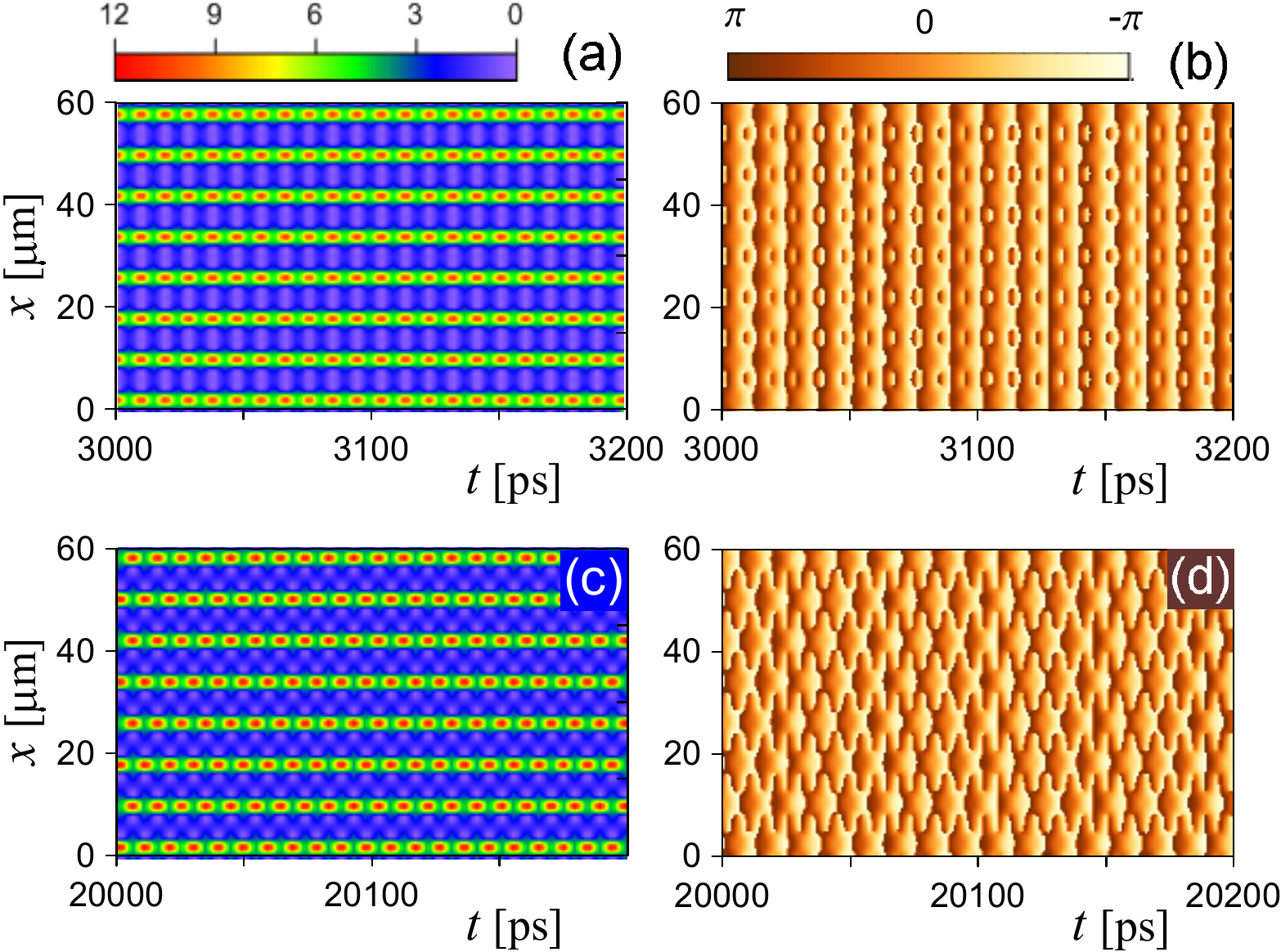}
\caption{ (Color online) Oscillation dynamics of the condensate calculated within the original model~(\ref{eq.polar}),(\ref{eq.reserv}). (a)  The $|\Psi|^2$ profile of the meta-stable oscillations for the potential contrasts $V_0 \left/ \hbar \right.=0.4$~{ps$^{-1}$}. (b) The phase profile of the dynamics shown in (a). (c) and (d) The same as in (a) and (b), respectively, after a long time ($t>10000$~ps). Other parameters: $P_0=18$~{$\mu$m$^{-2}$ps$^{-1}$}.}
\label{fig:Dynam_GL2}
\end{figure}

We also recalculated destabilization dynamics of the ground state for a stronger value of the potential contrast $V_0$ [Figs.~\ref{fig:Dynam_GL2} (a),(b)]. The temporal beating between the ground eigenstate ($\mu_{gr}$) and the excited symmetric one ($\mu_{sym}$) can be clearly observed, which is in a full agreement with the results of the simplified calculations [see Eq.~\eqref{eq:Omega_ZeroSt} and the Fig.~\ref{fig:Dynam_ZeroSt} (c) and (e)]. However, in accordance with the results of the numerical simulations of the original model~(\ref{eq.polar}),~(\ref{eq.reserv}), this dynamical state is meta-stable. It means that after a long time (more than several tens of nanoseconds) the dynamics is transformed spontaneously [Figs.~\ref{fig:Dynam_GL2} (c),(d)]. The analysis of the phase profile [Fig.~\ref{fig:Dynam_GL2} (d)]  shows that these hybrid oscillations appear between   ‘Zero-state’ and   ‘$\pi$-states’ of the condensate. Apparently, the selective gain saturation associated with the reservoir dynamics is responsible for this particular choice of the condensate eigenstates involved into the oscillations. We found that these types of oscillations are very robust and can develop even from the noisy initial conditions.
%
%

\section{Conclusion}

In this work we considered nonlinear dynamics of   coherent exciton-polaritons within the periodical potential embedded into a planar microresonator driven by a homogeneous incoherent pump. We restricted ourselves by weak-contrast lattices. 
Within this approach we developed a simplified mean-field model for three spatial harmonics and found analytical expressions for the relevant eigenstates of the condensate. 
Detailed theoretical study of the condensate dynamics in the vicinity of the threshold $P_{th}$ in the center and at the boundary of the Brillouin zone was carried out. In particular, we demonstrated   coherent persistent oscillations between different condensate eigenstates which are similar to those known for  Josephson junctions. The numerical results have been supported by the analytical analysis. Also   strong influence of the dissipative effects and the incoherent reservoir on the dynamics of  coherent polaritons has been discussed.

Apparently, for strong nonlinear system confined in lattices, the influence of the exciton-polariton condensate nonlinearity on band structure becomes important and occurs with the increase of pumping rate $P_0$. As for atomic systems confined in a periodic potential it is well known that the inter-particle interactions can lead to the formation of loops (so-called ``swallow tails'') both in the center and at the edge of the Brillouin zone \cite{Pethick03,Mueller02,Biao03}. Such loops in the energy band structure have been interpreted as the indicators of superfluid condensate properties~\cite{Biao03} and   hysteretic behavior of the condensate in respect to the variance of wave vector \cite{Mueller02}. More dramatic differences from the linear regime occur again in the center and at the edges of the Brillouin zone. Notably, such a system exhibits qualitatively new featrues of superfluid behavior of exciton-polaritons, affecting their energetic and dynamical stability~\cite{Biao03}. We hope to represent relevant results in the forthcoming paper.

\section*{ACKNOWLEDGMENTS}

This work was supported by RFBR Grants No 14-02-92604, 15-59-30406, 15-52-52001, 14-02-31443, No. 14-02-97503; by the Russian Ministry of Education and Science,
state task No 2014/13. O.A.E. and X.M. acknowledge financial support by the Deutsche Forschungsgemeinschaft (DFG project EG344/2-1) and the Thuringian Ministry for Education, Science and Culture (TMESC project B514-11027). The financial support from the EU project (FP7, PIRSES-GA-2013-612600) LIMACONA is acknowledged. A.P.A. acknowledges support from ``Dynasty'' Foundation.


\begin{thebibliography}{10}
\newcommand{\enquote}[1]{``#1''}
\expandafter\ifx\csname url\endcsname\relax
  \def\url#1{{#1}}\fi
\expandafter\ifx\csname urlprefix\endcsname\relax\def\urlprefix{}\fi

\bibitem{Stringari03}
M. Kramer, C. Menotti, L. Pitaevskii, and S. Stringari, Eur. Phys. J. D \textbf{27}, 247-261 (2003);  M. Lewenstein, A. Sanpera, V. Ahufinger, B. Damski, Aditi Sen De, and Ujjwal Sen, Adv. in Phys.  \textbf{56}, 243 (2007);  A. Kantian, J. Daley, P. T\"orm\"a, and P. Zoller, New J. of Physics \textbf{9}, 407 (2007).

\bibitem{IBloch2005}
I. Bloch, J. Phys. B. At. Mol. Opt. Phys. \textbf{38}, S629 (2005); O. Morsch, M. Oberthaler, Rev. of Mod. Phys. \textbf{78}, 179 (2006).

\bibitem{Drummod2011}
Q.Y. He, M.D. Reid, T.G. Vaughan, C. Gross, M. Oberthaler, and P.D. Drummond, Phys. Rev. Letts \textbf{106}, 120405 (2011); M. Hafezi, A.S. Sorensen, E. Demler, and M.D. Lukin, Phys. Rev. A \textbf{76}, 023613 (2007); D. Jaksch, P. Zoller, Annals of Physics \textbf{315}, 52 (2005); I. Bloch, Nature \textbf{453}, 1016 (2008)


\bibitem{Weisbuch92}
C.~Weisbuch, M.~Nishioka, A.~Ishikawa, and Y.~Arakawa, 
Phys. Rev. Lett. {\bf 69}, 3314 (1992).

\bibitem{Houdre94}
R.~Houdre, R.~Stanley, U.~Oesterle, M.~Ilegems, and C.~Weisbuch,
  Phys. Rev. B {\bf 49}, 16\,761 (1994).

\bibitem{Sanvitto2012}
D.~Sanvitto and V.~Timofeev, {\em Exciton Polaritons in Microcavities\/}
  (Springer Verlag, Berlin, 2012).

\bibitem{Kasprzak06}
J.~Kasprzak, M.~Richard, S.~Kundermann, A.~Baas, P.~Jeambrun, J.M. Keeling,
  F.M. Marchetti, M.H. Szymanska, R.~Andre, J.L. Staehli, V.~Savona, P.~B.
  Littlewood, B.~Deveaud, and L.S. Dang,
  Nature (London) {\bf 443}, 409 (2006).

\bibitem{Wouters07a}
M.~Wouters and I.~Carusotto,
Phys. Rev. Lett. {\bf 99}, 140402 (2007).


\bibitem{Deveaud12}
B.~Deveaud-Pl\'{e}dran, 
J. Opt.  Soc. Am. B {\bf 29}, 138 (2012).

\bibitem{Tredicucci96}
A.~Tredicucci, Y.~Chen, V.~Pellegrini, M.~B\"{o}rger, and F.~Bassani,
  Phys. Rev. A {\bf 54}, 3493 (1996).

\bibitem{Baas04}
A.~Baas, J.~P. Karr, H.~Eleuch, and E.~Giacobino, 
  Phys. Rev. A {\bf 69}, 023809 (2004).

\bibitem{Bajoni08}
D.~Bajoni, E.~Semenova, A.~Lemaitre, S.~Bouchoule, E.~Wertz, P.~Senellart,
  S.~Barbay, R.~Kuszelewicz, and J.~Bloch,
  Phys. Rev. Lett. {\bf 101}, 266402 (2008).

\bibitem{Ciuti00}
C.~Ciuti, P.~Schwendimann, B.~Deveaud, and A.~Quattropani, 
  Phys. Rev. B {\bf 62}, 4825(R)   (2000).

\bibitem{Savvidis00}
P.~G. Savvidis, J.~J. Baumberg, R.~M. Stevenson, M.~S. Skolnick, D.~M.
  Whittaker, and J.~S. Roberts,
  Phys. Rev. Lett. {\bf 84}, 1547 (2000).

\bibitem{Whittaker01}
D.~M. Whittaker, 
  Phys. Rev. B {\bf 63}, 193305 (2001).

\bibitem{Ciuti03}
C.~Ciuti, P.~Schwendimann, and A.~Quattropani,
 Semicond. Sci.  Technol. {\bf 18}, 279 (2003).

\bibitem{Wouters07}
M.~Wouters and I.~Carusotto,
Phys. Rev. B {\bf  75}, 075332 (2007).

\bibitem{Lagoudakis08}
K.~Lagoudakis, M.~Wouters, M.~Richard, A.~Baas, I.~Carusotto, R.~Andre,
  L.~Dang, and B.~Deveaud-Pledran,
  Nature Physics {\bf 4}, 706 (2008).

\bibitem{Lagoudakis09}
K.~G. Lagoudakis, T.~Ostatnick\'{y}, A.~V. Kavokin, Y.~G. Rubo, R.~Andr\'{e},
  and B.~Deveaud-Pl\'{e}dran,
  Science {\bf 326}, 974 (2009).

\bibitem{Amo11a}
A.~Amo, S.~Pigeon, D.~Sanvitto, V.~G. Sala, R.~Hivet, I.~Carusotto,
  F.~Pisanello, G.~Lem\'{e}nager, R.~Houdr\'{e}, E.~Giacobino, C.~Ciuti, and
  A.~Bramati, 
  Science {\bf 332}, 1167 (2011).

\bibitem{Flayac11}
H.~Flayac, D.~D. Solnyshkov, and G.~Malpuech,
  Phys. Rev. B {\bf  83}, 193305 (2011).

\bibitem{Hivet12}
R.~Hivet, H.~Flayac, D.D. Solnyshkov, D.~Tanese, T.~Boulier, D.~Andreoli,
  E.~Giacobino, J.~Bloch, A.~Bramati, G.~Malpuech, and A.~Amo,
  Nat. Phys. {\bf 8}, 724 (2012).

\bibitem{Solnyshkov12a}
D.D.~Solnyshkov, H.~Flayac, and G.~Malpuech, 
Phys. Rev. B {\bf 85}, 073105 (2012).

\bibitem{Flayac13}
H.~Flayac, D.~D. Solnyshkov, I.A.~Shelykh, and G.~Malpuech,
  Phys. Rev. Lett. {\bf 110}, 016404 (2013).

\bibitem{Egorov10a}
O.A. Egorov, A.V. Gorbach, F.~Lederer, and D.V. Skryabin,
  Phys. Rev. Lett. {\bf 105}, 073903 (2010).

\bibitem{Egorov10b}
O.A. Egorov, D.V. Skryabin, and F.~Lederer, \enquote{Polariton solitons due
  to saturation of the exciton-photon coupling}, Phys. Rev. B {\bf 82},
  165326 (2010).

\bibitem{Egorov11a}
O.A. Egorov, D.V. Skryabin, and F.~Lederer, 
Phys. Rev. B {\bf  84}, 165305 (2011).

\bibitem{Sich12a}
M.~Sich, D.~Krizhanovskii, M.S. Skolnick, A.V. Gorbach, R.~Hartley,
  D.~Skryabin, E.A.~Cerda-M\'{e}ndez, K.~Biermann, R.~Hey, and P.~Santos,
  Nat. Photonics {\bf 6}, 50 (2012).

\bibitem{Sich14a}
M.~Sich, F.~Fras, J.K.~Chana, M.~Skolnick, D.~Krizhanovskii, A.V.~Gorbach,
  R.~Hartley, D.V.~Skryabin, S.S.~Gavrilov, E.A.~Cerda-M\'{e}ndez, K. Biermann, R.~Hey,
  and P.V.~Santos,
  Phys. Rev. Lett. {\bf 112},  046403 (2014).

\bibitem{Wertz10}
E.~Wertz, L.~Ferrier, D.~Solnyshkov, R.~Johne, D.~Sanvitto, A.~Lemaitre,
  I.~Sagnes, R.~Grousson, A.~V. Kavokin, P.~Senellart, G.~Malpuech, and
  J.~Bloch, 
  Nat. Phys. {\bf 6}, 860 (2010).

\bibitem{Kaitouni06}
R.~I. Kaitouni, O.~E. Da\"{\i}f, A.~Baas, M.~Richard, T.~Paraiso, P.~Lugan,
  T.~Guillet, F.~Morier-Genoud, J.~D. Gani\`{e}re, J.~L. Staehli, V.~Savona,
  and B.~Deveaud,
  Phys. Rev. B {\bf 74}, 155311 (2006).

\bibitem{Balili07}
R.~Balili, V.~Hartwell, D.~Snoke, L.~Pfeiffer, and K.~West,
   Science {\bf 316}, 1007 (2007).

\bibitem{Lai07}
C.~Lai, N.~Kim, S.~Utsumomiya, G.~Roumpos, H.~Deng, M.~Fraser, T.~Byrnes,
  P.~Recher, N.~Kumada, T.~Fujisawa, and Y.~Yamamoto,
  Nature   {\bf 450}, 529 (2007).

\bibitem{Kim13}
N.~Y. Kim, K.~Kusudo, A.~Loffler, S.~Hofling, A.~Forchel, and Y.~Yamamoto,
  New J. Phys. {\bf 15}, 035032 (2013).

\bibitem{Lima06}
M.~M. de~Lima, M.~van~der Poel, P.~V. Santos, and J.~M. Hvam,
Phys. Rev. Lett. {\bf 97},  045501 (2006).

\bibitem{Amo10c}
A.~Amo, S.~Pigeon, C.~Adrados, R.~Houdr\'{e}, E.~Giacobino, C.~Ciuti, and
  A.~Bramati, 
  Phys. Rev. B {\bf 82}, 081301(R) (2010).

\bibitem{Staliunas06}
K.~Staliunas, R.~Herrero, Germ\'{a}n, and J.~de~Valc\'{a}rce,
Phys. Rev. E {\bf 73}, 065603 (2006).

\bibitem{Staliunas11}
K.~Staliunas,
  Phys. Rev. A {\bf 84}, 013626   (2011).

\bibitem{Cerda10}
E.~A. Cerda-M\'{e}ndez, D.~N. Krizhanovskii, M.~Wouters, R.~Bradley,
  K.~Biermann, K.~Guda, R.~Hey, P.~V. Santos, D.~Sarkar, and M.~S. Skolnick,
   Phys. Rev.  Lett. {\bf 105}, 116402 (2010).

\bibitem{Krizhanovskii13}
D.~N. Krizhanovskii, E.~A. Cerda-M\'{e}ndez, S.~Gavrilov, D.~Sarkar, K.~Guda,
  R.~Bradley, P.~V. Santos, R.~Hey, K.~Biermann, M.~Sich, F.~Fras, and M.~S.
  Skolnick,
  Phys. Rev. B {\bf 87}, 155423 (2013).

\bibitem{Ostrovskaya13}
E.~A. Ostrovskaya, J.~Abdullaev, M.~D. Fraser, A.~S. Desyatnikov, and Y.~S.
  Kivshar, 
  Phys. Rev. Lett. {\bf 110}, 170407 (2013).

\bibitem{Egorov07d}
O.~A. Egorov and F.~Lederer,
  Phys. Rev. A {\bf 76}, 053816 (2007).

\bibitem{Egorov10c}
O.~A. Egorov, F.~Lederer, and K.~Staliunas,
Phys. Rev. A {\bf 82}, 43830 (2010).

\bibitem{Smirnov14}
L.~A. Smirnov, D.~A. Smirnova, E.~A. Ostrovskaya, and Y.~S. Kivshar,
  Phys. Rev. B {\bf 89}, 235310 (2014).

\bibitem{Wouters08}
M.~Wouters,
Phys. Rev. B {\bf 77}, 121302(R) (2008).

\bibitem{Borgh10}
M.~O. Borgh, J. Keeling, N.~G. Berloff,
Phys. Rev. B {\bf 81}, 235302 (2010).

\bibitem{Shelykh08}
I.~A. Shelykh, D.~D. Solnyshkov, G. Pavlovic, and G. Malpuech,
 Phys. Rev. B {\bf 78}, 041302(R) (2008).

\bibitem{Demirchyan14}
S.~S. Demirchyan, I.~Yu. Chestnov, A.~P. Alodjants, M.~M. Glazov, and A.~V. Kavokin,  
Phys. Rev. Lett. {\bf 112}, 196403 (2014).

\bibitem{Liew14}
T.~C.~H. Liew, Y.~G. Rubo, and A.~V. Kavokin,
Phys. Rev. B {\bf 90}, 245309 (2014).

\bibitem{Dominici14}
L. Dominici, D. Colas, S. Donati, J.~P. Restrepo Cuartas, M. De Giorgi, D. Ballarini, G. Guirales, J.~C. Lуpez Carreсo, A. Bramati, G. Gigli, E. del Valle, F.~P. Laussy, and D. Sanvitto, 
Phys. Rev. Lett. {\bf 113}, 226401 (2014).


\bibitem{Pethick03}
M. Machholm, C.~J. Pethick, H. Smith, 
Phys. Rev. A \textbf{67}, 053613 (2003).

\bibitem{Mueller02}
E.~J. Mueller,
Phys. Rev. A {\bf 66}, 063603 (2002)

\bibitem{Biao03}
Biao Wu, Qian Niu, 
New J. of Phys. \textbf{5}, 104 (2003).

\bibitem{Lagoudakis10}
K.~G. Lagoudakis, B.~Pietka, M.~Wouters, R.~Andr\'{e}, and B.~Deveaud-Pl\'{e}dran, 
Phys. Rev. Lett. {\bf 105}, 120403 (2010).

\bibitem{Jensen82}
S.M. Jensen, 
IEEE Journal of Quantum Electronics \textbf{18}, 1580 (1982).

\bibitem{Abramowitz72}
M. Abramowitz and I. A. Stegun, \textit{Handbook of Mathematical Functions with Formulas, Graphs, and Mathematical Tables} (National Bureau of Standards Applied Mathematics Series, Washington DC, 1972).
\end{thebibliography}

\end{document}